\documentclass[aps,prc,twocolumn]{revtex4-2}
\usepackage{amsmath,amssymb}
\usepackage{graphicx}
\usepackage{float}
\usepackage{soul}
\usepackage{xcolor}

\newcommand{\be}{\begin{equation}}
\newcommand{\ee}{\end{equation}}
\newcommand{\ba}{\begin{eqnarray}}
\newcommand{\ea}{\end{eqnarray}}
\newcommand{\ban}{\begin{eqnarray*}}
\newcommand{\ean}{\end{eqnarray*}}
\newcommand \nn {\nonumber}

\begin{document}

\title{Why magnetic-field effects are difficult to observe \\ in relativistic heavy-ion collisions}

\author{Stanis\l aw Mr\' owczy\' nski\footnote{e-mail: stanislaw.mrowczynski@ncbj.gov.pl} 
and Maciej Pylak\footnote{e-mail: maciej.pylak@ncbj.gov.pl}}

\affiliation{National Centre for Nuclear Research, ul. Pasteura 7, 02-093 Warsaw, Poland}

\date{November 26, 2026}

\begin{abstract}

Since a strong magnetic field is generated in relativistic heavy-ion collisions, novel phenomena have been predicted to occur in the quark-gluon plasma produced in these collisions. However, these predictions have not yet been convincingly confirmed experimentally. We reanalyze the generation and evolution of the magnetic field which plays a key role in the phenomena. Using the method of one-sided Fourier transformation, we reconcile the description in frequency and wave-vector space, which is required to properly account for the electromagnetic properties of the quark-gluon plasma, with the initial-value nature of the problem arising from the creation of the plasma medium at the moment of collision. We show that the currents induced in the quark-gluon plasma, which sustain the magnetic field, develop with a delay and are consequently less effective than usually assumed. Moreover, because the plasma initially consists predominantly of electrically neutral gluons, while charged quarks emerge at a later stage, the induced currents are further delayed. Consequently, a magnitude of the field present in the quark-gluon plasma is significantly lower than earlier estimates suggested. This may explain why magnetic-field effects are difficult to observe in relativistic heavy-ion collisions.

\end{abstract}

\maketitle

\section{Introduction}
\label{sec-into}

A magnetic field $B$ generated in relativistic heavy-ion collisions at RHIC and the LHC can reach a magnitude of the order of $10^{19}$~Gauss, making it one of the strongest magnetic fields found in nature; see the review \cite{Tuchin:2013ie}. This value corresponds to a field strength $eB$ of the order of $m_\pi^2$, where $e$ is the elementary electric charge and $m_\pi$ is the pion mass, which sets a characteristic scale of the strong interaction. Thus, one encounters the unusual situation in which the strength of the electromagnetic interaction becomes comparable to that of the strong interaction. One therefore expects a nontrivial interplay between electromagnetic and strong interactions; indeed, several effects arising from this interplay have been predicted. These include a shift in the temperature of the deconfinement phase transition in the presence of a magnetic field, the chiral magnetic effect, modifications of the directed collective flow of charged particles produced in relativistic heavy-ion collisions, and a splitting of the spin polarization of $\Lambda$ and $\bar{\Lambda}$ particles; see the review articles \cite{Tuchin:2013ie,Kharzeev:2012ph,Fukushima:2018grm,Huang:2015oca}. With the exception of a small modification of the directed collective flow \cite{STAR:2023bzh,STAR:2023jdd}, these predictions have not yet been confirmed experimentally; see \cite{Wang:2022eoo,STAR:2025vhs,Shen:2025unr}. Since the magnitude and lifetime of the magnetic field are crucial for these phenomena, the generation of magnetic fields in relativistic heavy-ion collisions requires an in-depth analysis.

In the first calculations \cite{Kharzeev:2007jp,Skokov:2009qp,Bzdak:2011yy,Voronyuk:2011jd}, the magnetic field was assumed to be generated by the electric charges of the colliding nuclei, moving at nearly the speed of light along straight-line trajectories, as well as by the charges of particles produced in the collisions. The field was found to reach the huge magnitude mentioned above but was estimated to decay by several orders of magnitude within 1~fm/$c$. It was soon realized that the field could be sustained for a longer time by electric currents induced in the quark-gluon plasma produced in the collisions \cite{Tuchin:2010vs,Deng:2012pc,McLerran:2013hla,Gursoy:2014aka,Zakharov:2014dia,Roy:2015kma,Li:2016tel,Huang:2022qdn}. This effect is governed by the electrical conductivity of the quark-gluon plasma, which was assumed to be frequency independent, {\it i.e.} static, in these studies. Later, the magnetic field was calculated using the full dielectric tensor of the quark-gluon plasma, which depends on both frequency and wave vector \cite{Grayson:2022asf}. In this case, the conductivity is also frequency and wave-vector dependent. Although the analysis of \cite{Grayson:2022asf} is the most advanced to date, it has one important deficiency: the plasma medium is assumed to be present even before the collision, whereas in reality, it is created only after the nuclei collide. Addressing this deficiency is the main motivation for our work.

To properly describe the electromagnetic properties of the quark-gluon plasma, one must work in frequency and wave-vector space $(\omega,{\bf k})$. However, the creation of the medium at the moment of the collision constitutes an initial-value problem, which is naturally formulated in time and position space $(t,{\bf r})$. At first glance, the two approaches appear incompatible since the transformation from time to frequency requires a Fourier transform with an integration over time from minus to plus infinity. This difficulty can be resolved by using the one-sided Fourier transform. If the initial condition is specified at $t = 0$, the integration over time is performed from zero to infinity. Of crucial importance is the fact that the medium response functions, in particular the dielectric tensor, are the same for both two-sided and one-sided Fourier transforms. This follows from the fact that the response functions satisfy the causality condition.

Since the plasma medium is created only at the moment of the collision, the magnetic field remains equal to its vacuum value until that time. The effect of currents induced in the medium is delayed, and for a longer time, the magnetic field is as in vacuum. An additional delay occurs because, initially, the medium is dominated by gluons, which carry no electric charge. We take this effect into account by assuming that the magnetic field remains equal to its vacuum value not only until the moment of the collision, $t=0$, but until some later time $t_0 > 0$. A delay in the onset of induced currents has been considered in a different way in \cite{Huang:2022qdn,Yan:2021zjc,Wang:2021oqq}.

This paper is mostly devoted to presenting a method based on the one-sided Fourier transform and comparing its results with those obtained using the usual two-sided transform. To keep the calculations as analytical as possible, we make two important simplifications: we use dielectric functions in the long-wavelength limit, {\it i.e.}, independent of the wave vector and we treat the electric charges of the colliding nuclei as point-like. In Sec.~\ref{sec-finite-size} we discuss the latter approximation and compute the magnetic field generated by the nuclei with a Gaussian distribution of electric charges. However, we apply only the two-sided Fourier transformation in Sec.~\ref{sec-finite-size}. Detailed results for both the magnetic and electric fields, obtained using the one-sided Fourier transformation, a realistic charge distribution within the nuclei and a more accurate dielectric tensor, will be presented in a subsequent publication.

We use natural units, $\hbar=c=1$, combined with the CGS system of electromagnetic units.

\section{Computational method}

Let us imagine that two relativistic heavy nuclei (both of charge $Ze$) collide at $t=0$ and produce a quark-gluon plasma. We are interested in the magnetic field in the plasma. For $t < 0$ the magnetic field is generated by the approaching nuclei. For $t > 0$ the field is still generated by the electrically charged remnants of the nuclei moving away from each other. However, due to Faraday's law, the quark currents are induced in the plasma, and the magnetic field is modified. 

\subsection{General formula}

We obtain the magnetic field by solving Maxwell's equations, which in the CGS system of units  are
\be
\label{Maxwell-eqs-x}
\begin{aligned}
&\nabla \cdot {\bf D}(t, {\bf r}) = 4\pi \rho_{\rm ext} (t, {\bf r}) ,
\\
&\nabla \cdot {\bf B}(t, {\bf r}) = 0 , 
\\
&\nabla \times {\bf E}(t, {\bf r}) = 
- {\partial {\bf B}(t, {\bf r}) \over \partial t} ,
\\
&\nabla \times {\bf B}(t, {\bf r}) =
4 \pi {\bf j}_{\rm ext}(t, {\bf r}) +{\partial {\bf D}(t, {\bf r}) \over \partial t} ,
\end{aligned}
\ee
where $\rho_{\rm ext} (t, {\bf r})$ and ${\bf j}_{\rm ext}(t, {\bf r})$ are the external densities of charge and current, ${\bf E}(t, {\bf r})$ and ${\bf B}(t, {\bf r})$ are the electric field and magnetic induction, while ${\bf D}(t, {\bf r})$ is the electric induction or electric displacement field. 

We note the a magnetic field ${\bf H}$ is absent in the Maxwell equations (\ref{Maxwell-eqs-x}). As discussed in detail in \cite{Rukhadze-Silin-1961}, the magnetic field ${\bf H}$, which is related to ${\bf B}$ via the magnetic permeability tensor $\mu^{ij}$, is not really needed, as the electric permeability $\varepsilon^{ij}$ is sufficient to encompass both electric and magnetic properties of a medium. 

To solve the equations (\ref{Maxwell-eqs-x}) we apply the one-sided Fourier transformation defined as
\be
f(\omega,{\bf k}) \equiv \int_0^\infty dt \int d^3r
e^{i(\omega t - {\bf k}\cdot {\bf r})}
f(t,{\bf r}) ,
\ee
with the inverse transformation equal
\be
\label{inv-one-sided-Fourier}
f(t,{\bf r}) = \int_{-\infty +i\sigma}^{\infty +i\sigma}
{d\omega \over 2\pi} \int {d^3k \over (2\pi)^3}
e^{-i(\omega t - {\bf k}\cdot {\bf r})} f(\omega,{\bf k}) ,
\ee
where the real parameter $\sigma > 0$ is chosen in such a way that the integral over $\omega$ is taken along a straight line in the complex $\omega-$plane, parallel to the real axis, above all singularities of $f(\omega,{\bf k})$. 

The transformed Maxwell equations are
\be
\label{Maxwell-eqs-k}
\begin{aligned}
&i {\bf k} \cdot \hat\varepsilon (\omega,{\bf k}) \, {\bf E}(\omega,{\bf k}) 
= 4\pi \rho_{\rm ext} (\omega,{\bf k}) ,
\\
&{\bf k}\cdot {\bf B}(\omega,{\bf k}) = 0 , 
\\
&i {\bf k} \times {\bf E}(t, {\bf r}) 
= i \omega {\bf B}(\omega,{\bf k}) + {\bf B}_0({\bf k}) ,
\\
&i {\bf k} \times {\bf B}(\omega,{\bf k}) =
4 \pi {\bf j}_{\rm ext}(\omega,{\bf k}) 
- i \omega \hat\varepsilon (\omega,{\bf k}) \, {\bf E}(\omega,{\bf k}) 
\\
&~~~~~~~~~~~~~~~~~~~~~~~~~~~~~~~~~~~~~~~~~~~~~~~~~~
- {\bf D}_0({\bf k}) ,
\end{aligned}
\ee
where ${\bf B}_0({\bf k})$ and ${\bf D}_0({\bf k})$ are the initial values at $t=0$ and the electric induction is expressed as 
\be
\label{D-vs-E}
D^i(\omega, {\bf k}) = \varepsilon^{ij}(\omega, {\bf k}) \,
E^j(\omega, {\bf k}) ,
\ee
with $ \varepsilon^{ij}(\omega, {\bf k})$ being the dielectric tensor. 

Using the Maxwell equations (\ref{Maxwell-eqs-k}), the electric field $E^i(\omega, {\bf k})$ is found as
\ba
\nn
&& E^i (\omega, {\bf k}) = -i (\Sigma^{-1})^{ij}(\omega,{\bf k})
\\ \label{E-field-k}
&&~~\times
\Big[ 4 \pi \omega j_{\rm ext}^j(\omega,{\bf k})
+ \epsilon^{jkl} k^k B_0^l ({\bf k})
- \omega D_0^j ({\bf k}) \Big] ,
\ea
where 
\be
\label{matrix-sigma}
\Sigma^{ij}(\omega,{\bf k}) \equiv
- {\bf k}^2 \delta^{ij} + k^ik^j
+ \omega^2 \varepsilon^{ij}(\omega,{\bf k}) .
\ee

Assuming that the medium is isotropic, the dielectric tensor is of the form
\be
 \varepsilon^{ij} (\omega,{\bf k}) =  \varepsilon_L (\omega,{\bf k}) \frac{k^i k^j}{{\bf k}^2}
 +  \varepsilon_T (\omega,{\bf k})\Big( \delta^{ij} - \frac{k^i k^j}{{\bf k}^2}\Big) ,
 \ee
 and 
\ba
\nn
\Sigma^{ij}(\omega,{\bf k}) &=& \omega^2 \varepsilon_L (\omega,{\bf k}) \frac{k^i k^j}{{\bf k}^2}
\\ \label{matrix-sigma-iso}
&& 
\big(\omega^2\varepsilon_T (\omega,{\bf k}) -{\bf k}^2 \big) \Big( \delta^{ij} - \frac{k^i k^j}{{\bf k}^2}\Big) .
\ea
Consequently,
\ba
\nn
(\Sigma^{-1})^{ij}(\omega,{\bf k}) &=& \frac{1}{\omega^2\varepsilon_L (\omega,{\bf k})} \frac{k^i k^j}{{\bf k}^2}
\\
&+&  \frac{1}{\omega^2\varepsilon_T (\omega,{\bf k}) - {\bf k}^2} \Big( \delta^{ij} - \frac{k^i k^j}{{\bf k}^2}\Big) ,
\ea
and the electric field (\ref{E-field-k}) equals
\begin{widetext}
\be
\label{E-field-k-2}
E^i (\omega, {\bf k}) = -i \Big[ \frac{1}{\omega^2\varepsilon_L (\omega,{\bf k})} \frac{k^i k^j}{{\bf k}^2}
 +  \frac{1}{\omega^2\varepsilon_T (\omega,{\bf k}) - {\bf k}^2} \Big( \delta^{ij} - \frac{k^i k^j}{{\bf k}^2}\Big)\Big]
\Big[ 4 \pi \omega j_{\rm ext}^j(\omega,{\bf k})
+ \epsilon^{jkl} k^k B_0^l ({\bf k})
- \omega D_0^j ({\bf k}) \Big] .
\ee
Since, according to the Faraday's law,
\be
B^i(\omega,{\bf k}) = \frac{1}{\omega}\epsilon^{ijk}k^j E^k(\omega,{\bf k}) 
+ \frac{i}{\omega}  B_0^i({\bf k}) , 
\ee
the magnetic field is found as
\be
\label{B-field-k-final}
{\bf B} (\omega, {\bf k}) = - \frac{i}{\omega^2\varepsilon_T (\omega,{\bf k}) - {\bf k}^2}  
\Big[ {\bf k} \times \Big(4 \pi {\bf j}_{\rm ext}(\omega,{\bf k}) - {\bf D}_0 ({\bf k})\Big) 
- \frac{{\bf k}^2}{\omega} \, {\bf B}_0 ({\bf k})\Big]  + \frac{i}{\omega} \, {\bf B}_0 ({\bf k}) ,
\ee
which is the main result of this section.
\end{widetext}
We note that when ${\bf D}_0 ({\bf k}) =  {\bf B}_0 ({\bf k}) = 0$, we get 
\be
\label{B-field-k-final-static}
{\bf B}(\omega,{\bf k}) = -    
\frac{4 \pi i}{\omega^2  \varepsilon_T (\omega,{\bf k})  - {\bf k}^2} \,
 {\bf k} \times {\bf j}_{\rm ext}(\omega,{\bf k}) ,
\ee
which looks as the solution of the Maxwell equations obtained by means of the usual two-sided Fourier transformation. However, one should remember that although the dielectric function is the same for the one- and two-sided transformations, 
the meanings of ${\bf B}(\omega,{\bf k})$ and ${\bf j}_{\rm ext}(\omega,{\bf k})$ are different, as the inverse one- and two-sided transformations are different. 

In the derivation above, the initial condition is imposed at $t=0$. However, by modifying the method as described in Appendix \ref{app-initial-t0}, the initial condition can be imposed at any time. We will make use of this possibility in Sec.~\ref{sec-results}.

\subsubsection{External current}

As noted in the introduction, the colliding nuclei are treated here as point-like charges $Ze$ moving with constant velocity ${\bf v}$ along the trajectory ${\bf r}(t) = {\bf r}_0 + {\bf v}t$. The charge and current densities are
\ba
\label{charge-density}
\rho_{\rm ext}(t,{\bf r}) &=& Ze \, \delta^{(3)}({\bf r} - {\bf r}_0 - {\bf v}t) ,
\\[2mm]
\label{current}
{\bf j}_{\rm ext}(t,{\bf r}) &=& Ze \, {\bf v} \delta^{(3)}({\bf r} - {\bf r}_0 - {\bf v}t) .
\ea

The one-sided Fourier transform of the current (\ref{current}) is computed as
\ba
\nn
{\bf j}_{\rm ext}(\omega,{\bf k}) &=&  \int_0^\infty dt \int d^3r
e^{i(\omega t - {\bf k}\cdot {\bf r})} \, {\bf j}_{\rm ext}(t,{\bf r}) 
\\ \nn
&=& Ze \,{\bf v} \,  e^{-i {\bf k} \cdot  {\bf r}_0} 
\int_0^\infty dt \, e^{i(\omega  - {\bf k}\cdot {\bf v}) t} 
\\
&=& Ze \,{\bf v} \,  e^{-i {\bf k} \cdot  {\bf r}_0}  
\frac{e^{i(\omega  - {\bf k}\cdot {\bf v}) t}}{i(\omega  - {\bf k}\cdot {\bf v})}\bigg|_0^\infty .
\ea
To make the limit $t \to \infty$ meaningful, we replace $\omega \to \omega +i0^+$ and then,
\be
\label{current-one-sided-Fourier}
{\bf j}_{\rm ext}(\omega,{\bf k}) = i \frac{Ze \, {\bf v} \, e^{-i {\bf k} \cdot  {\bf r}_0} }
{\omega  - {\bf k}\cdot {\bf v} +i0^+} .
\ee

\subsection{Initial values}

To use the formula (\ref{B-field-k-final}), the initial fields ${\bf D}_0 ({\bf k})$ and ${\bf B}_0 ({\bf k})$ have to be specified. The initial fields are the fields generated by the incoming nuclei at $t = 0$. Since for the $t < 0$ there is no plasma ${\bf D}_0 ({\bf k}) = {\bf E}_0 ({\bf k})$.  In vacuum $\varepsilon^{ij} (\omega,{\bf k}) =  \delta^{ij}$ and the initial fields are found using the two-sided Fourier transformation. Modifying Eqs.~(\ref{E-field-k-2}) and (\ref{B-field-k-final-static}), we have
\ba
\label{E0}
{\bf E}_0 ({\bf k}) &\equiv& {\bf E}(t=0,{\bf k}) 
\\ \nn
&=& - 4 \pi i \int_{-\infty}^\infty \frac{d\omega}{2\pi}
 \Big[\frac{\omega {\bf j}_{\rm ext}(\omega,{\bf k}) }{\omega^2 - {\bf k}^2}  
 - \frac{{\bf k} \big({\bf k} \cdot {\bf j}_{\rm ext}(\omega,{\bf k})\big)}{\omega(\omega^2 - {\bf k}^2)}\Big] ,
\\[4mm]
\label{B0}
{\bf B}_0({\bf k}) &\equiv& {\bf B}(t=0,{\bf k}) = - 4 \pi i \int_{-\infty}^\infty \frac{d\omega}{2\pi} \,
\frac{ {\bf k} \times {\bf j}_{\rm ext}(\omega,{\bf k})}{\omega^2 - {\bf k}^2} .
\ea
Due to the continuity of the electromagnetic current, we have the relation $\omega \rho_{\rm ext} (\omega,{\bf k}) = {\bf k} \cdot {\bf j}_{\rm ext} (\omega,{\bf k})$ and ${\bf E}_0 ({\bf k})$ can be written as
\be
\label{E0_1}
{\bf E}_0 ({\bf k}) = - 4 \pi i \int_{-\infty}^\infty \frac{d\omega}{2\pi} \,
 \frac{\omega {\bf j}_{\rm ext}(\omega,{\bf k}) - {\bf k} \rho_{\rm ext} (\omega,{\bf k})}{\omega^2 - {\bf k}^2}  .
\ee

Since the two-sided Fourier transforms of the formulas (\ref{charge-density}) and (\ref{current}) are
\ba
\label{charge-den-k}
\rho_{\rm ext}(\omega,{\bf k})  &=& 
2\pi \delta(\omega - {\bf k} \cdot {\bf v})  \, Ze \, e^{-i {\bf k} \cdot  {\bf r}_0},
\\[2mm]
\label{current-k}
{\bf j}_{\rm ext}(\omega,{\bf k})  &=& 
2\pi \delta(\omega - {\bf k} \cdot {\bf v})  \, Ze {\bf v} \, e^{-i {\bf k} \cdot  {\bf r}_0},
\ea
the initial fields equal
\ba
\label{E0-point}
{\bf E}_0 ({\bf k}) &=& - 4 \pi i \, Ze \, 
 \frac{({\bf k} \cdot  {\bf v}) {\bf v} - {\bf k}}{({\bf k} \cdot  {\bf v})^2 - {\bf k}^2}  
\, e^{-i {\bf k} \cdot  {\bf r}_0} ,
\\[4mm]
\label{B0-point}
{\bf B}_0({\bf k}) &=& - 4 \pi i  \, Ze \,
\frac{{\bf k} \times {\bf v}}{({\bf k} \cdot  {\bf v})^2 - {\bf k}^2} 
 \, e^{-i {\bf k} \cdot  {\bf r}_0}.
\ea

One can check the formulas (\ref{E-field-k}) and (\ref{B-field-k-final}) observing that if the nuclei move in a vacuum or in a medium which does not change from $t=-\infty$ to $t=\infty$, the one-sided and two-sided Fourier transformations should give the same result for $t>0$.

\subsection{Dielectric functions}

The dielectric functions $\varepsilon_L(\omega,{\bf k})$ and $\varepsilon_T(\omega,{\bf k})$, which enter the expressions for the electric and magnetic fields, Eqs.~(\ref{E-field-k-2}) and (\ref{B-field-k-final}), should properly describe the electromagnetic properties of an isotropic quark-gluon plasma. Since the electromagnetic field generated in relativistic heavy-ion collisions varies rapidly in time at a fixed position, the dielectric functions should be reliable at high frequencies $\omega$. If the electric charge of a colliding nucleus is modeled as a point-like charge and we consider the field away from the location of the point charge, the field is not strongly nonuniform at a fixed time. It is therefore sufficient to use the dielectric functions valid in the long-wavelength limit, {\it i.e.}, for small ${\bf k}$. The well-known dielectric functions of a collisionless plasma \cite{LP81} that satisfy these requirements are
\be
\label{dielectric-fun-colless}
 \varepsilon_{L,T} (\omega,{\bf k})  = 1 - \frac{\omega_p^2}{\omega^2} ,
\ee
where $\omega_p$ is the plasma frequency. The dielectric functions (\ref{dielectric-fun-colless}) ignore dissipative effects, and in particular the corresponding conductivity diverges. Taking into account collisions among the plasma constituents in the relaxation-time approximation, the dielectric functions become \cite{LP81} 
\be
\label{dielectric-fun-coll}
 \varepsilon_{L,T} (\omega,{\bf k})  = 1 - \frac{\omega_p^2}{\omega (\omega + i\nu)} ,
\ee
where $\nu$ is the collision frequency, or equivalently, the inverse relaxation time. This is the transverse dielectric function that we use in our calculations.

Since the magnetic field generated in relativistic heavy-ion collisions is usually parameterized in terms of the plasma conductivity, let us relate the dielectric function (\ref{dielectric-fun-coll}) to the conductivity. The conductivity tensor $\sigma^{ij}(\omega,{\bf k})$ is defined as 
\be
j_{\rm ind}^i(\omega,{\bf k}) = \sigma^{ij}(\omega,{\bf k}) \, E^j(\omega,{\bf k}) ,
\ee
where ${\bf j}_{\rm ind}$ is the induced current. Using the Maxwell equations (\ref{Maxwell-eqs-k}), one finds
\be
\label{relation}
\sigma^{ij}(\omega,{\bf k}) = \frac{i\omega}{4\pi} \big(\delta^{ij} - \varepsilon^{ij}(\omega,{\bf k})\big) ,
\ee
where the factor $1/(4\pi)$ arises from our use of the CGS system of electromagnetic units. It is absent in the rationalized system of units. For an isotropic medium, Eq.~(\ref{relation}) with the dielectric functions (\ref{dielectric-fun-coll}) gives
\be
\label{conductivity-TL}
\sigma_{L,T}(\omega,{\bf k}) = \frac{i\omega}{4\pi} \big(1-\varepsilon_{L,T}(\omega,{\bf k})\big)
= \frac{i\omega_p^2}{4\pi(\omega+i\nu)} ,
\ee
which reduces to the static conductivity 
\be
\label{sigma-stat}
\sigma_{\rm stat} = \frac{\omega_p^2}{4 \pi \nu}
\ee
in the limit $\omega\rightarrow 0$.

We note that the conductivity (\ref{conductivity-TL}) is equivalent to that given by Eqs.~(40) and (41) in \cite{Grayson:2022asf} which is called the Drude model there.

\section{Computation}

We consider a collision of two nuclei moving along the straight-line trajectories ${\bf r}_+(t) =(b/2, 0, vt)$ and ${\bf r}_-(t) =(-b/2, 0, -vt)$, where $b$ is the impact parameter, chosen to lie along the $x$ axis. Since electrodynamics is a linear theory, the electromagnetic field generated by the two nuclei is simply the sum of the fields generated by the individual nuclei. We therefore first compute the magnetic field generated by a single nucleus moving along the trajectory ${\bf r}(t) =(0, 0, vt)$. Because the problem is cylindrically symmetric, it is sufficient to compute only the $y$ component of the magnetic field in the plane $y=0$. The $x$ and $y$ components of the field at an arbitrary spatial point can then be obtained by rotating the coordinate system around the $z$ axis. Shifting the coordinate system by $b/2$ along the axis $x$ we get the magnetic field generated by the nucleus moving along the trajectory ${\bf r}_+(t) =(b/2, 0, vt)$. Replacing ${\bf r}_+(t)$ with ${\bf r}_-(t)$ yields the field generated by the second nucleus, and the total field is obtained by summing the two contributions.

Substituting the external current (\ref{current-one-sided-Fourier}), the initial fields (\ref{E0-point}) and (\ref{B0-point}), as well as the dielectric function (\ref{dielectric-fun-coll}) into Eq.~(\ref{B-field-k-final}), we obtain a complete analytic expression for ${\bf B}(\omega,{\bf k})$. To determine the field ${\bf B}(t,{\bf r})$, the one-sided Fourier transformation must be inverted according to Eq.~(\ref{inv-one-sided-Fourier}). Expressing the wave vector in cylindrical coordinates $(k_T,\phi,k_z)$, such that $k_x=k_T\cos\phi$ and $k_y=k_T\sin\phi$, the inversion requires the evaluation of a fourfold integral over $\omega$, $k_T$, $\phi$, and $k_z$. The frequency integral can be evaluated analytically using the residue theorem. 

The integrand in Eq.~(\ref{B-field-k-final}) has four simple poles as a function of $\omega$, and thus there are four contributions to the frequency integral and each contribution has a clear physical meaning. The dispersion equation $\omega^2\varepsilon_T(\omega,{\bf k})-{\bf k}^2=0$, with the dielectric function (\ref{dielectric-fun-coll}), is a third-order equation and therefore yields three poles corresponding to transverse electromagnetic plasmons excited by the external electric current in the quark-gluon plasma. The current (\ref{current-one-sided-Fourier}) gives rise to a pole at $\omega={\bf k}\cdot{\bf v}$, which represents the coupling of the electric charge generating the current and moving with velocity ${\bf v}$ to the magnetic field with phase velocity $\omega/{\bf k}$. We also note that $\omega=0$ is not a pole of the integrand (\ref{B-field-k-final}) with the dielectric function (\ref{dielectric-fun-coll}) because the two terms associated with the initial magnetic field, which diverge as $\omega \to 0$, cancel each other in this limit. 

Since we compare the results obtained using the one-sided Fourier transformation with those obtained using the usual two-sided Fourier transformation, we also perform analogous calculations for the latter case, which is, however, significantly simpler. Instead of the four terms in Eq.~(\ref{B-field-k-final}), Eq.~(\ref{B-field-k-final-static}) contains only one term, and the integration over $\omega$ is trivial because of the delta function $\delta(\omega-{\bf k}\cdot{\bf v})$ appearing in the charge density (\ref{charge-den-k}) and current (\ref{current-k}). The magnetic field obtained using the two-sided Fourier transformation corresponds to the contribution from the pole $\omega={\bf k}\cdot{\bf v}$ in the one-sided Fourier transformation.

Similarly to the frequency integral, the integral over the azimuthal angle can be evaluated analytically as well. It yields a Bessel function of the first kind $J_1\big((x-b/2)k_T\big)$. The remaining integrals over $k_z$ and $k_T$ are then evaluated numerically. In case of $k_z$ we apply the method of Fast Fourier Transformation and we exploit the cylindrical symmetry, computing the integral over $k_T$ {\it via} a Discrete Hankel Transform \cite{Ronen:2006}, as explained in Appendix~\ref{app-Hankel}.

\section{Parameter values}

The magnetic field generated by colliding nuclei depends on several parameters. If the nuclei are modeled as point-like charges and the effects of the plasma medium are absent, the magnetic field is determined by the charges of the nuclei, their velocity, and the impact parameter. We consider the collision of two gold nuclei at the top RHIC energy $\sqrt{s_{NN}} = 200$~GeV. Thus, each nucleus carries a charge $Ze$, with $Z=79$, and the Lorentz factor of each nucleus is $\gamma=107.5$ in the laboratory frame, in which the magnetic field is calculated. The impact parameter is chosen to be 6~fm.

The electromagnetic properties of the quark-gluon plasma are encoded in the dielectric functions (\ref{dielectric-fun-coll}), which depend on two parameters: the plasma frequency $\omega_p$ and the collision frequency $\nu$. To get a rough estimate of these parameters, we assume that we are dealing with an equilibrium quark-gluon plasma at the temperature $T$. Since  the square of the plasma frequency of the collisionless electron-positron plasma is $\omega_p^2 = 4\pi \alpha T^2/9$ (in CGS units), where $\alpha = 1/137$ is the fine-structure constant, see Eq.~(6.69) of \cite{Le-Bellac-2000}, one estimates the electromagnetic plasma frequency of the quark-gluon plasma as
\be
\label{plasma-frequency-quarks}
\omega_p^2 = \frac{20 \pi}{27} \alpha T^2 ,
\ee
where the contributions of the $u$, $\bar{u}$, $d$, and $\bar{d}$ quarks, with electric charges $\pm 2e/3$ and $\mp e/3$, respectively, have been taken into account. For $T = 400$~MeV, we obtain $\omega_p \approx 52$~MeV which is used in our further calculations performed at fixed $T$. Since the plasma cools as it expands, the plasma temperature should be considered time-dependent. We plan to address this effect in our next publication.

The collision frequency $\nu$ can be estimated \cite{Mrowczynski:1988xu}, using the parton density in the weakly interacting quark-gluon plasma at the temperature $T$ and the parton-parton transport cross section. The result is 
\be
\label{nu-kinetic}
\nu = \frac{10 (9N_f + 16) \zeta(3)}{17\pi} \,\alpha_s^2 \ln (\alpha_s^{-1}) \, T,
\ee
where $\zeta(z)$ is the Riemann zeta function and $\zeta (3) \approx 1.20$. For $N_f = 2$, $T = 400$~MeV and the coupling constant from the interval $\alpha_s \in [0.1, 0.3]$, Eq.~(\ref{nu-kinetic}) gives
\be
\label{nu-est-1}
\nu \in [70, 331] ~~[{\rm MeV}] .
\ee

\begin{table}[b]
\caption{\label{tab-parameters} The parameters}
\center
\begin{tabular}{c  c  c}
\hline
\hline
\noalign{\smallskip}
 ~~ parameter~~ & ~~~~ value ~~~~ & ~~ unit ~~ \\
\hline
\hline
\noalign{\smallskip}
$Z$ & 79 & \\
\hline
\noalign{\smallskip}
$\alpha = e^2$ & 1/137 & \\
\hline
\noalign{\smallskip}
$\gamma$ & 107.5 &  \\
\hline
\noalign{\smallskip}
$b$ & 6 & fm \\
\hline
\noalign{\smallskip}
$t_0$ & 0, 0.1, 0.3 & fm/$c$ \\
\hline
\noalign{\smallskip}
$\omega_p$ & 52 & MeV \\
\hline
\noalign{\smallskip}
$\nu$ & 100, 200, 400 & MeV \\
\hline
\noalign{\smallskip}
$R_{\rm rms}$ & 5.3 & fm \\
\hline
\hline
\end{tabular}
\end{table}

The collision frequency $\nu$ can also be found in such a way that the static conductivity (\ref{sigma-stat}) reproduces the lattice result \cite{Ding:2010ga} which for the two-flavor plasma is
\be
\sigma_{\rm stat} =\frac{1}{4\pi} (5.6 \pm 2.8) \, \frac{T}{T_c} ~~[{\rm MeV}],
\ee
where $T_c = 165$~MeV. The result is actually obtained for $T = 1.45 \, T_c \approx 240$~MeV but we will use it for the higher temperature $T = 400$~MeV. Then, $\sigma_{\rm stat} = (13.6 \pm 6.8)/(4\pi)$~MeV. Using $\omega_p \approx 52$~MeV, the formula (\ref{sigma-stat}) gives 
\be
\label{nu-est-2}
\nu \in [133, 398] ~~[{\rm MeV}].
\ee
One sees that the estimates (\ref{nu-est-1}) and (\ref{nu-est-2}) overlap rather well. Our calculations are performed for $\nu = 100, 200$ and 400 MeV, which correspond to the relaxation time $2, 1$ and 0.5 fm/$c$, respectively.

\begin{figure}[t]
\centering
\includegraphics[width=8.5cm]{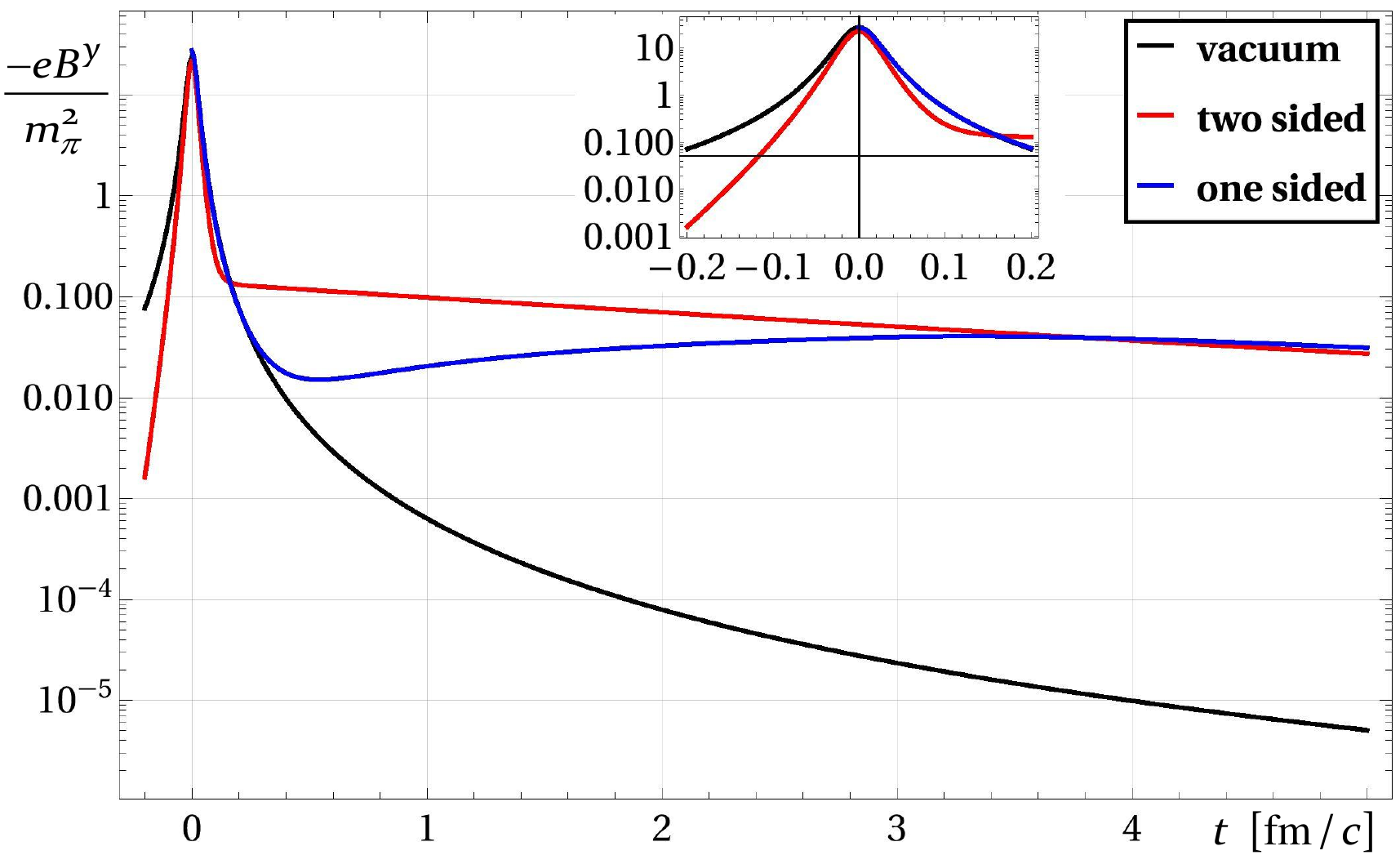}
\includegraphics[width=8.5cm]{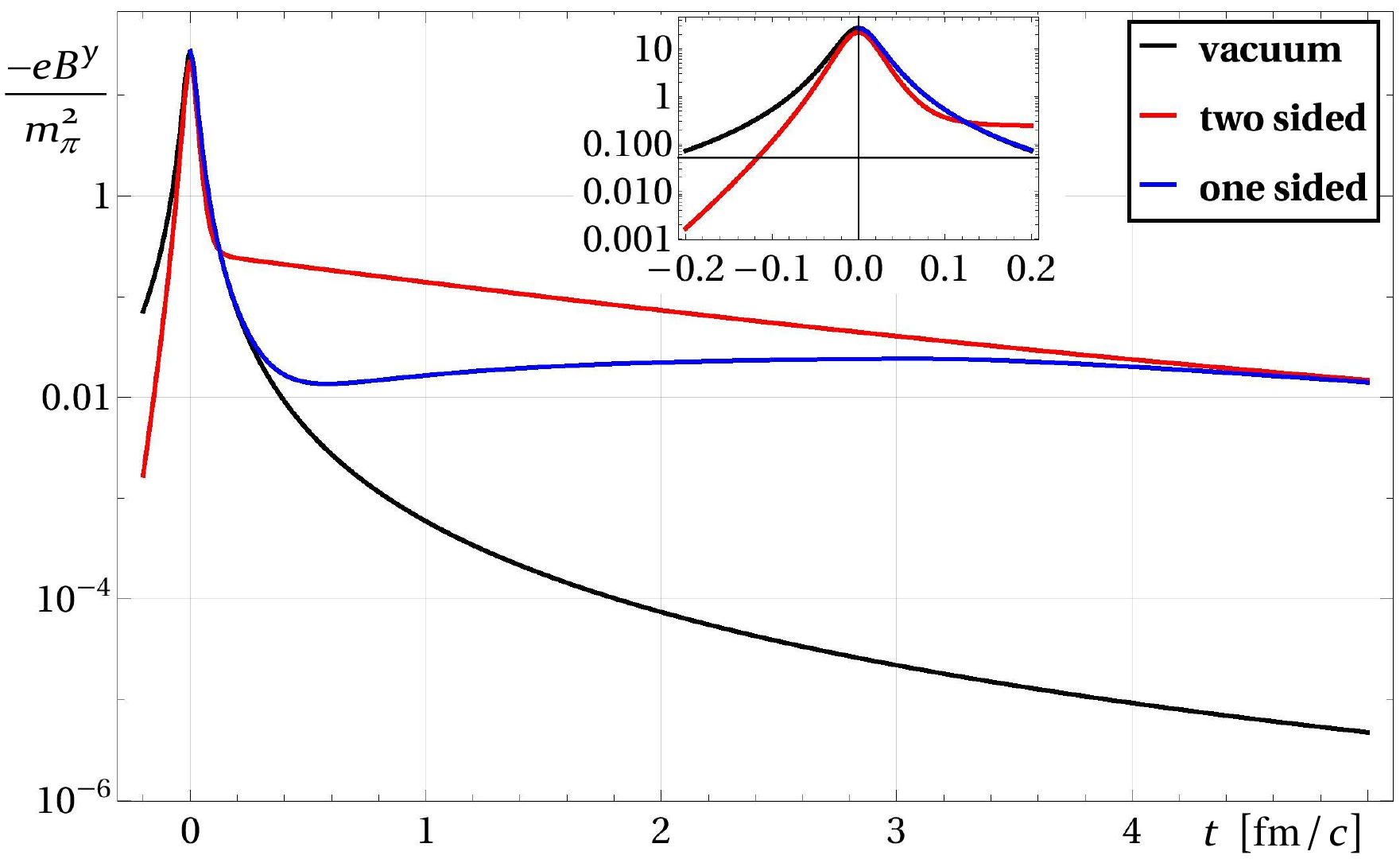}
\includegraphics[width=8.5cm]{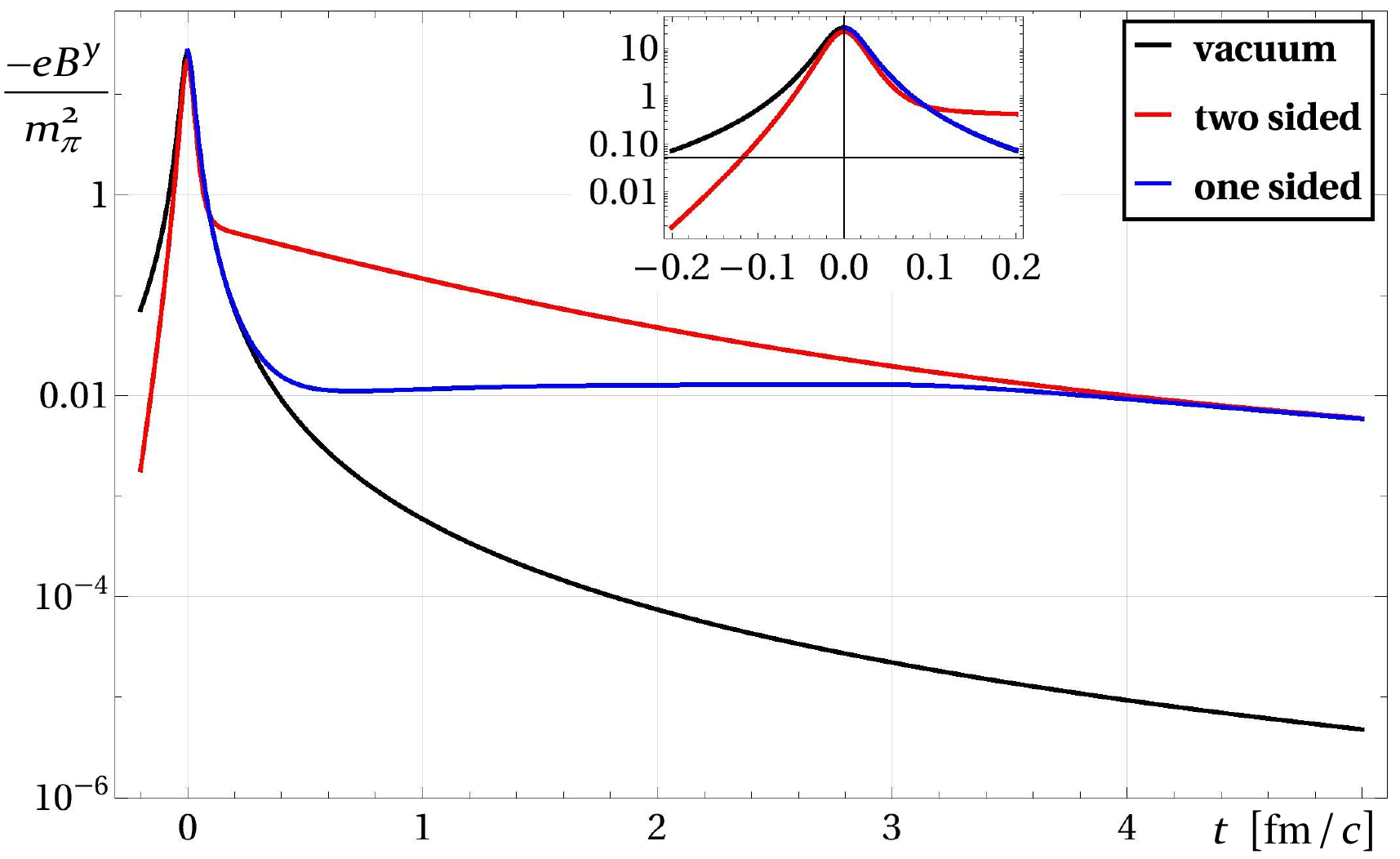}
\vspace{-5mm}
\caption{The magnetic field generated in vacuum (black) and in the quark-gluon plasma computed with the one-sided (blue) and two-sided (red) Fourier transformation for the collision frequency $\nu = 100$ MeV (upper panel), 200 MeV (middle panel) and 400 MeV (lower panel). }
\label{Fig-B-nu}
\end{figure}

\section{Results}
\label{sec-results}

\begin{figure}[t]
\centering
\includegraphics[width=8.5cm]{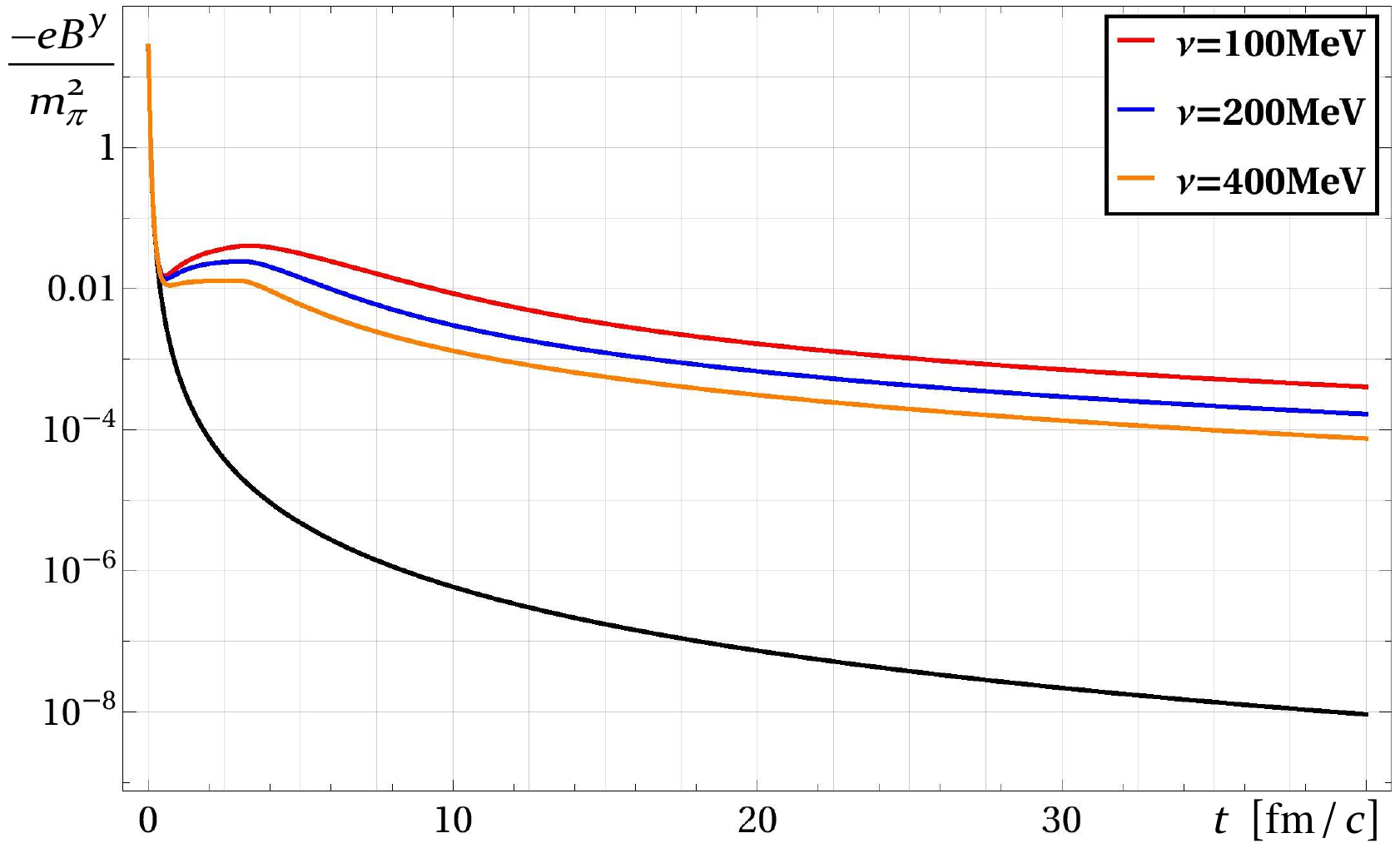}
\includegraphics[width=8.5cm]{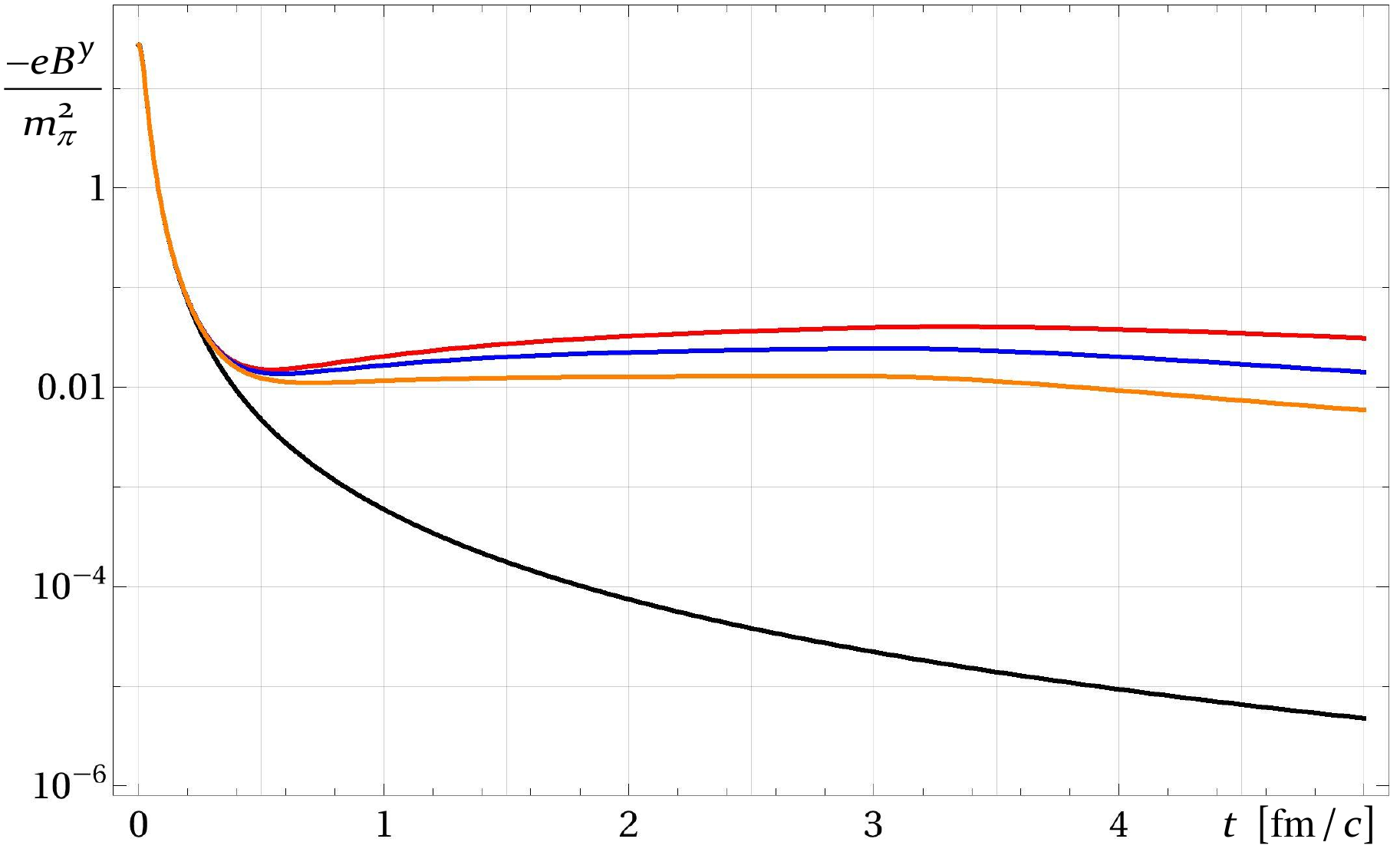}
\vspace{-3mm}
\caption{The magnetic field generated in vacuum (black) and in the quark-gluon plasma computed with the one-sided Fourier transformation for the collision frequency $\nu = 100$ MeV (red), 200 MeV (blue) and 400 MeV (orange). The upper and lower panel show the magnetic field in the long and short time intervals $0 < t < 40~{\rm fm}/c$ and $0 < t < 5~{\rm fm}/c$, respectively. }
\label{Fig-B-long-t}
\end{figure}

\begin{figure}[t]
\centering
\includegraphics[width=8.5cm]{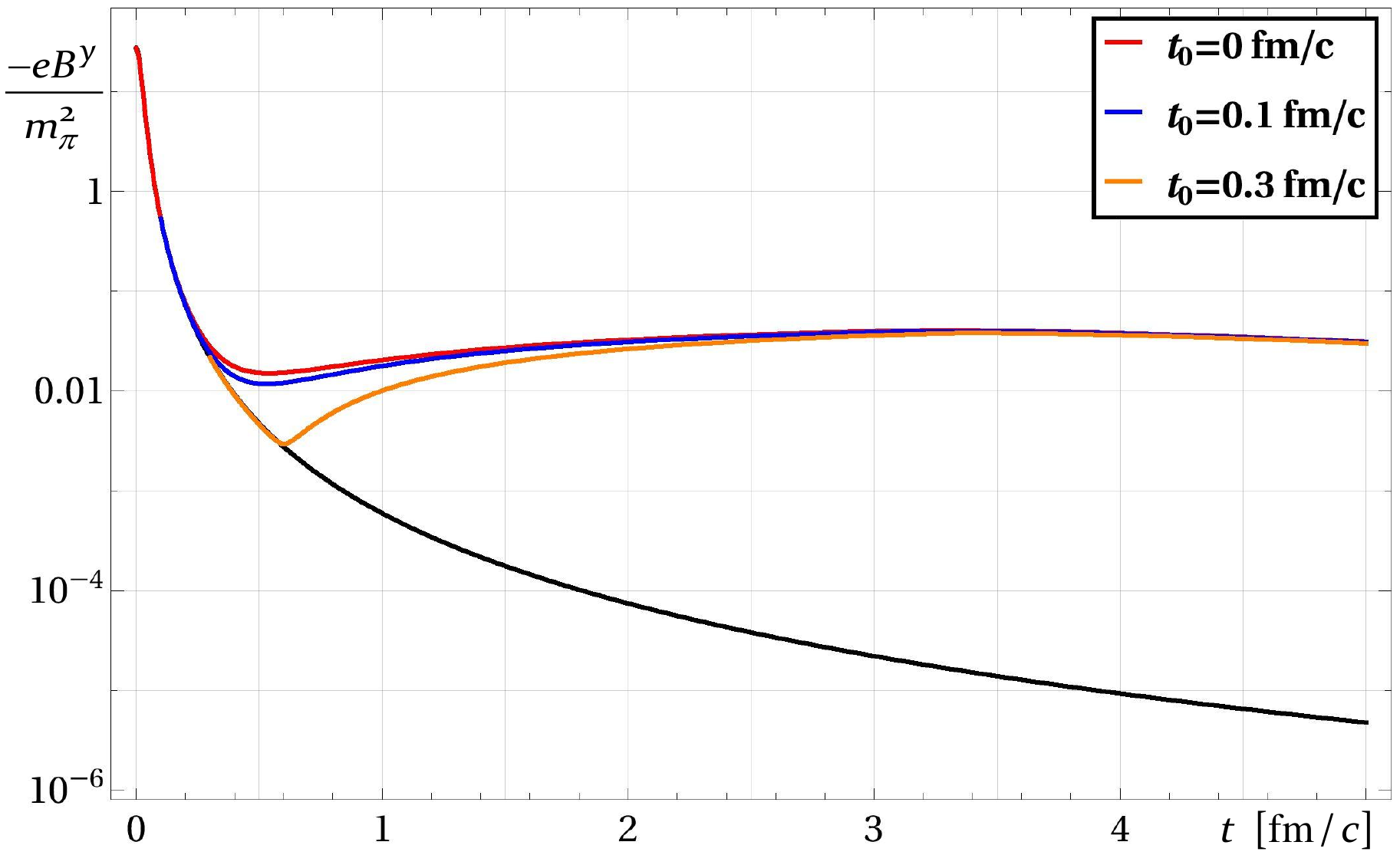}
\includegraphics[width=8.6cm]{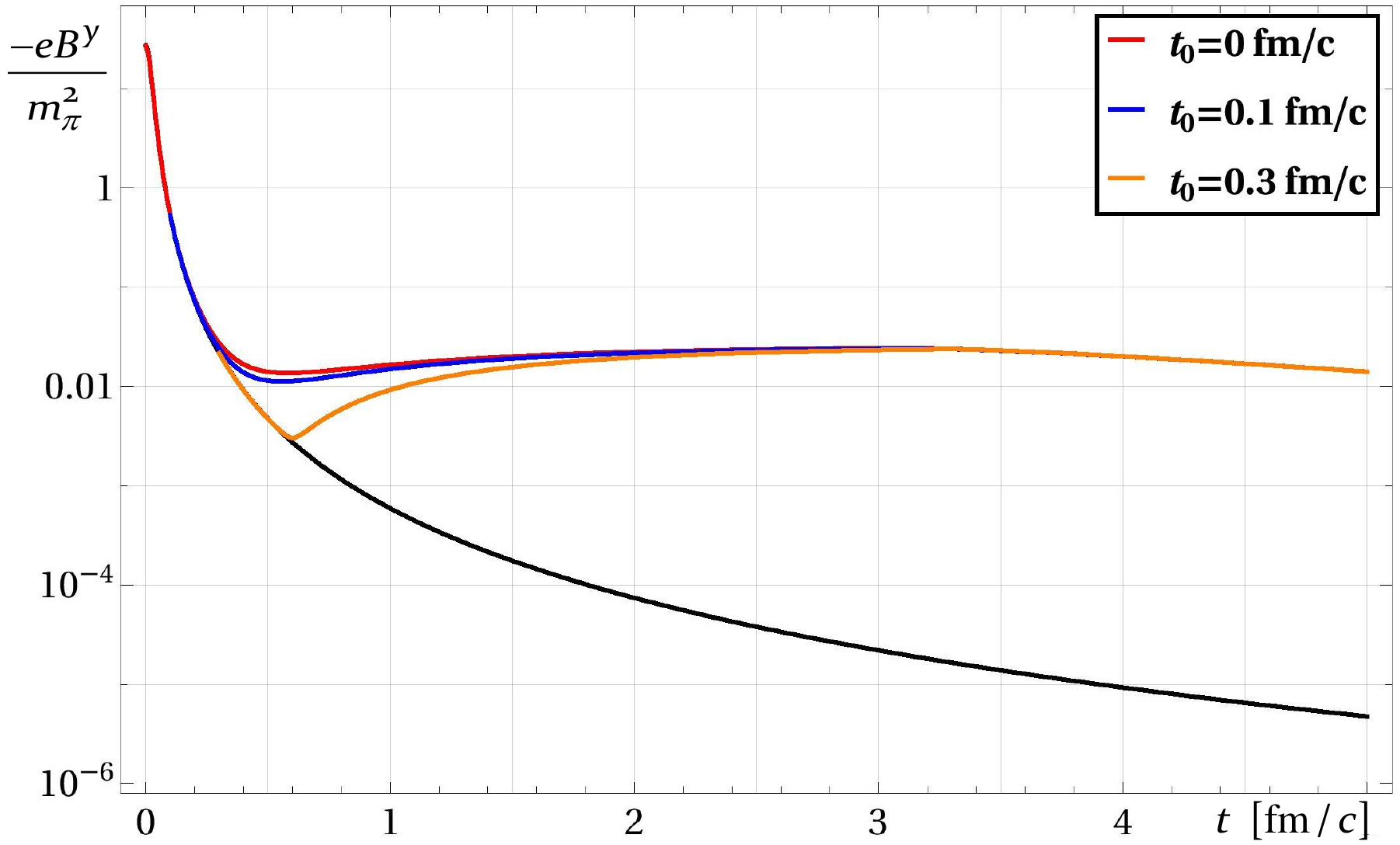}
\includegraphics[width=8.5cm]{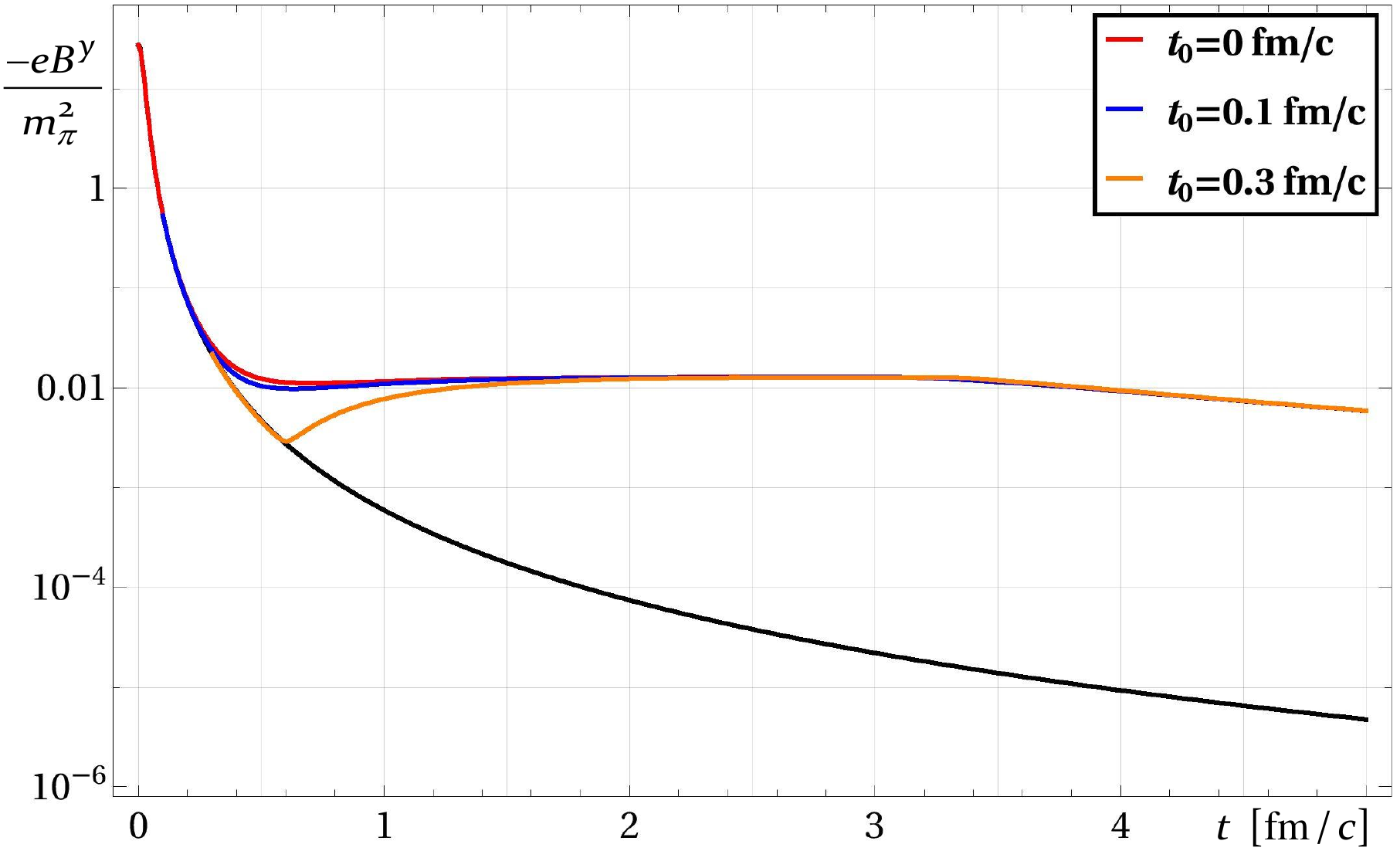}
\vspace{-3mm}
\caption{The magnetic field generated in vacuum (black) and in the quark-gluon plasma computed with the one-sided Fourier transformation with the initial condition at $t_0=0$ (red), $t_0=0.1$~fm/$c$ (blue), and $t_0=0.3$~fm/$c$ (orange) for the collision frequency $\nu = 100$ MeV (upper panel), 200 MeV (middle panel) and 400 MeV (lower panel).}
\label{Fig-B-t0}
\end{figure}

Using the parameters collected in Table~\ref{tab-parameters}, we have computed the magnetic field as a function of time at the point $x=y=z=0$ that is halfway between the colliding nuclei, where the field is directed along the $y$-axis. The results are shown in Figs.~\ref{Fig-B-nu}-\ref{Fig-B-finite-size-medium}.

In Fig.~\ref{Fig-B-nu} we compare the magnetic field generated in vacuum with that in the quark-gluon plasma, calculated using the one- and two-sided Fourier transformations, for the collision frequencies $\nu = 100$, 200, and 400 MeV, which correspond to the electric conductivities $\sigma_{\rm stat} = 27/(4\pi),~13.5/(4\pi)$, and $6.8/(4\pi)$~MeV (in CGS units). When the field is generated in vacuum, it is symmetric under time reversal, $t \to -t$. This symmetry is lost when a medium extending from the remote past to the remote future is assumed and the two-sided Fourier transformation is used to evaluate the field. As the nuclei approach each other and the magnetic field grows, the currents induced in the plasma oppose its growth, causing the field to increase more slowly than in vacuum. Conversely, as the nuclei move apart and the magnetic field decreases, the induced currents oppose its decrease, causing the field to decay more slowly than in vacuum. This behavior is a manifestation of Lenz's law. We observe that the medium's effect in sustaining the magnetic field is very significant. When the medium is created at $t=0$ and the one-sided Fourier transformation is used to evaluate the magnetic field, it is exactly the same as in vacuum for $t < 0$. For $t \lesssim 0.5\;{\rm fm}/c$, it remains approximately equal to the vacuum field. The field-sustaining effect becomes apparent only for $t \gtrsim 0.5\;{\rm fm}/c$, and even then it is noticeably weaker than that obtained using the two-sided Fourier transformation.

The magnetic field calculated using the one-sided Fourier transformation and shown in Fig.~\ref{Fig-B-nu} exhibits an unexpected behavior: it does not decrease for $t \gtrsim 1.5\;{\rm fm}/c$, but instead grows. However, this is merely a transient effect. As shown in the upper panel of Fig.~\ref{Fig-B-long-t}, the field decays at later times. Nevertheless, it should be noted that the quark-gluon plasma hadronizes at a time of about $10$~fm$/c$, and thus Fig.~\ref{Fig-B-long-t} has no physical significance beyond this time. The lower panel of Fig.~\ref{Fig-B-long-t} shows how the magnetic field depends on the collision frequency in the time domain relevant to relativistic heavy-ion collisions. As can be seen, the stronger the electrical conductivity of the plasma, the stronger the magnetic field.

Using the one-sided Fourier transform allows the initial condition to be imposed not only at $t_0=0$ but at any arbitrary moment in time, as discussed in Appendix A. Since the quark-gluon plasma is initially dominated by gluons -- which carry no electric charge -- the medium effect must be delayed until electrically charged quarks appear. Until that point, the magnetic field behaves as it would in a vacuum. Fig.~\ref{Fig-B-t0} shows the magnetic field calculated using the one-sided Fourier transform for the initial vacuum conditions set at $t_0=0,~0.1$, and 0.3~fm/$c$, for the collision frequencies $\nu = 100$, 200, and 400 MeV. As one observes, the dependence on $t_0$ is very weak as long as $t_0$ is less than the time required for the field-sustaining effect to emerge.

Our calculations are performed at the fixed temperature $T=400$~MeV, whereas the plasma actually cools as it expands. Consequently, the plasma conductivity is expected to decrease over time, meaning that the magnetic field we obtain is overestimated at later times. We plan to investigate this effect in detail in a future publication.

\section{Effect of nucleus finite size}
\label{sec-finite-size}

\begin{figure}[t]
\centering
\includegraphics[width=8.6cm]{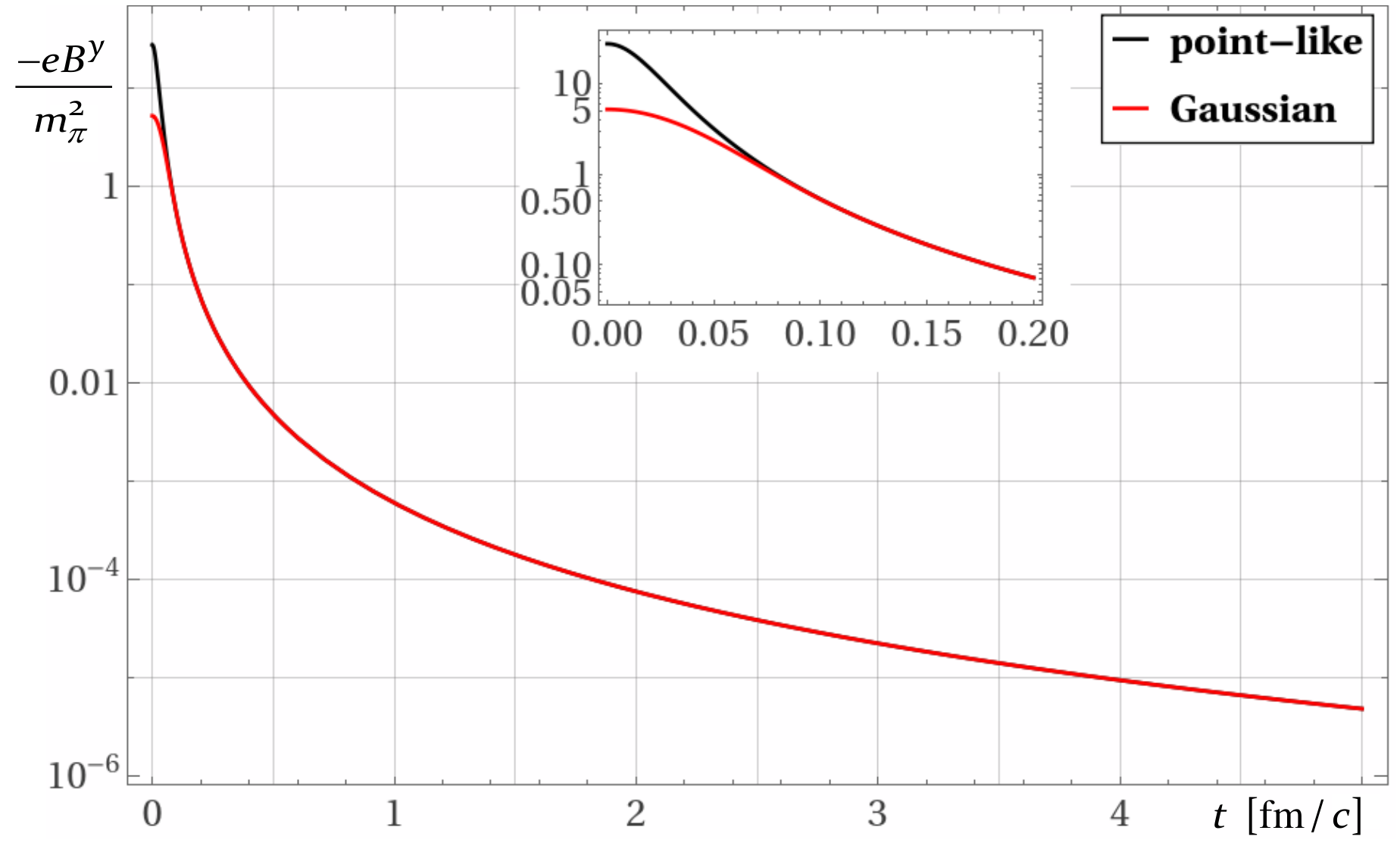}
\vspace{-3mm}
\caption{The magnetic field generated in vacuum by point-like nuclei (orange) and by nuclei of Gaussian distribution of electric charges (blue).}
\label{Fig-finite-size}
\end{figure}

Our results have been obtained by treating the electric charges of the colliding nuclei as point-like. This is obviously a rather drastic approximation when the magnetic field is evaluated at a distance from the nucleus smaller than its radius. However, the nuclei we consider are ultrarelativistic and strongly Lorentz-contracted. Therefore, neglecting the finite nuclear size is expected to affect the magnetic field only during the short time interval in which a nucleus passes through the point where the field is observed.

To investigate the effects of finite nuclear size and test this expectation, we adopt, as in Ref.~\cite{Grayson:2022asf}, a Gaussian charge distribution characterized by the parameter $R$, which is related to the root-mean-square nuclear radius $R_{\rm rms}$ as   $R = \sqrt{2/3} \, R_{\rm rms}$, and the total charge $Ze$. This choice is motivated by the fact that the Fourier transform can then be calculated analytically, whereas the more realistic Woods-Saxon distribution would require numerical computation. With this choice, the external charge and current densities are given not by Eqs.~(\ref{charge-density}) and~(\ref{current}), but, for the geometry considered here, by
\ba
\label{charge-density-finite}
\rho_{\rm ext}(t,{\bf r}) &=& \frac{Ze \gamma}{\pi^{3/2}R^3} \, 
e^{- \frac{1}{R^2}\big[(x \mp b/2)^2 + y^2 +\gamma^2(z \mp vt)^2\big]},
\\[2mm]
\label{current-finite}
{\bf j}_{\rm ext}(t,{\bf r}) &=& {\bf v} \, \rho_{\rm ext}(t,{\bf r}) ,
\ea
where $\gamma \equiv (1 - {\bf v}^2)^{-1/2}$ is the Lorentz gamma factor and ${\bf v} = (0,0,\pm v)$. The two-sided Fourier transform of the charge density (\ref{charge-density-finite}) is
\be
\label{Gauss-charge}
\rho_{\rm ext}(\omega,{\bf k}) = Z e \, e^{- \frac{1}{4}(k_x^2 + k_y^2 +k_z^2/\gamma^2) R^2} 
e^{\mp \frac{i}{2} k_x b} \,  2\pi \delta(\omega \mp k_z v) ,
\ee
where the upper and lower signs correspond to the two oppositely moving nuclei. We used this expression to calculate the magnetic field generated by the colliding gold nuclei $^{197}{\rm Au}$ in vacuum at the origin of our coordinate system for $\gamma = 107.5$,  $b=6$ fm and $R=4.3$ fm which corresponds to $R_{\rm rms}=5.3$ fm \cite{DeJager:1974liz}. The result is compared with that obtained for point-like nuclei in Fig.~\ref{Fig-finite-size}.

As seen in the figure, the magnetic field calculated for point-like nuclei is significantly stronger than that obtained with the finite-size charge distribution only at times shorter than 0.1 fm/$c$, corresponding to the time interval during which the Lorentz-contracted nuclei pass through the point where the field is evaluated. This is consistent with our expectation that the effects of finite nuclear size are confined to a very short time interval. However, when the magnetic field is generated in the plasma medium the situation is more complex, as we explain below.  

\begin{figure}[t]
\centering
\includegraphics[width=8.5cm]{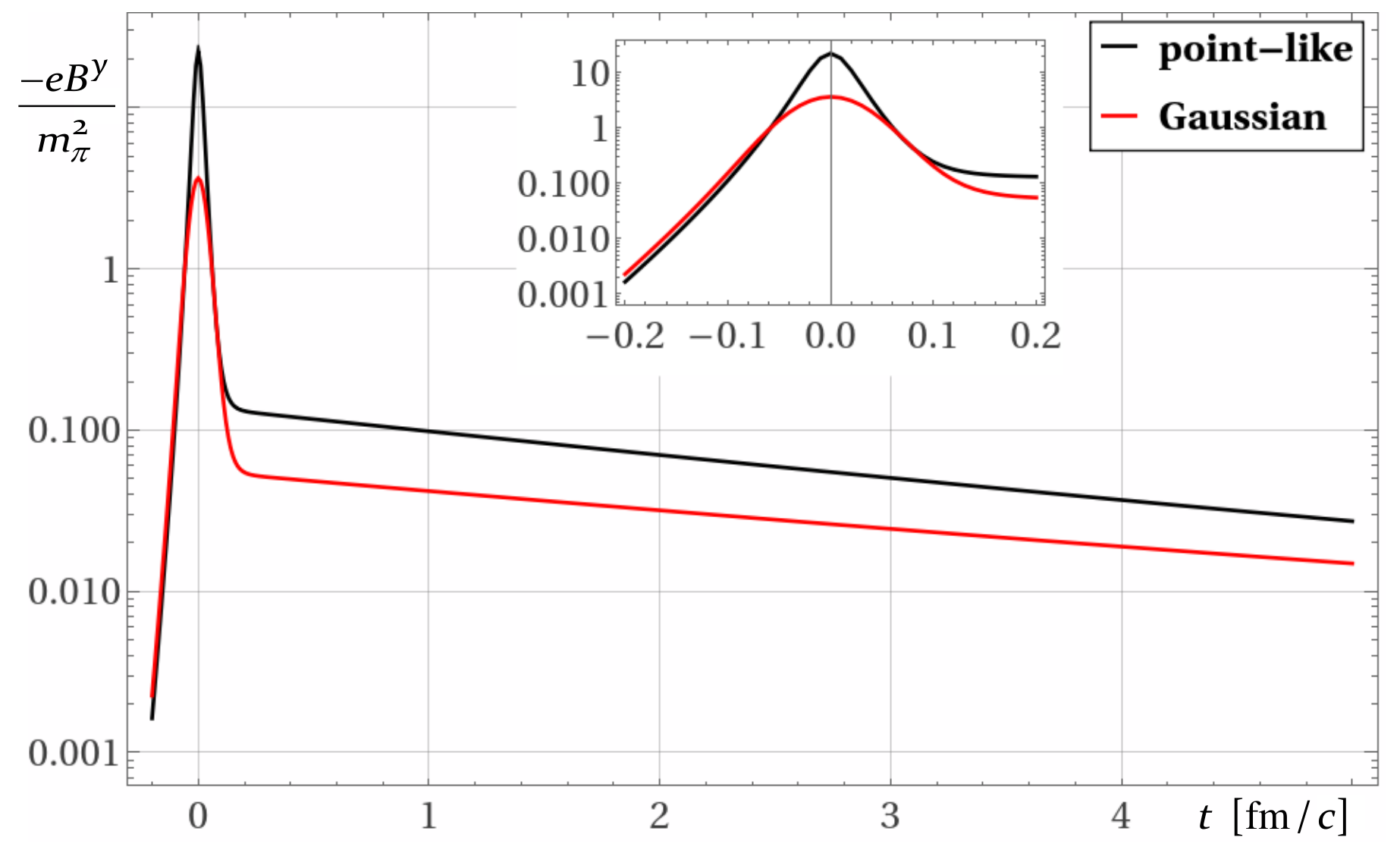}
\includegraphics[width=8.5cm]{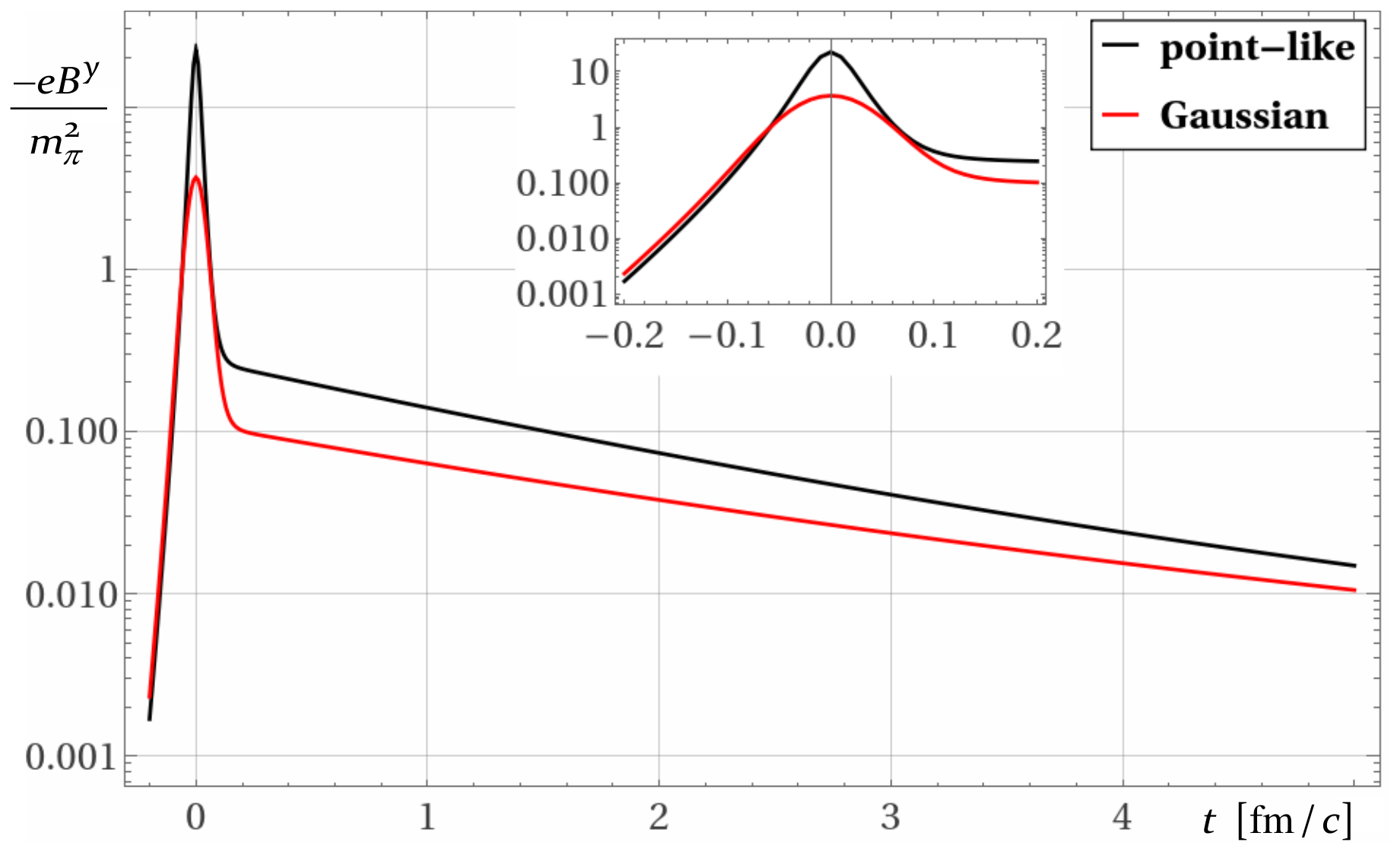}
\includegraphics[width=8.5cm]{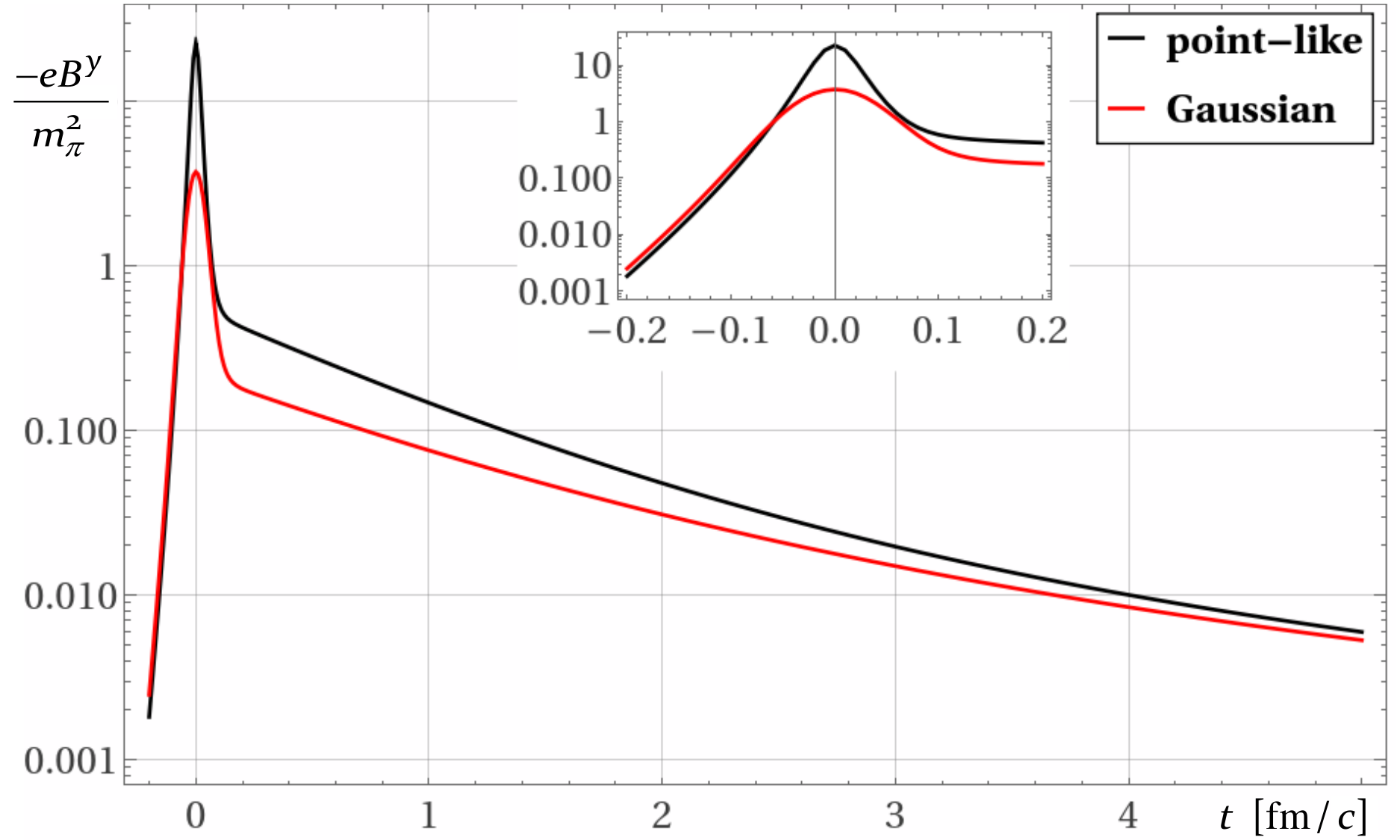}
\vspace{-3mm}
\caption{The magnetic field generated in vacuum (black) and in the quark-gluon plasma computed with the two-sided Fourier transformation for the collision frequency $\nu = 100$ MeV (upper panel), 200 MeV (middle panel) and 400 MeV (lower panel).}
\label{Fig-B-finite-size-medium}
\end{figure}

Fig.~\ref{Fig-B-finite-size-medium} shows the magnetic field calculated for the Gaussian charge distribution (\ref{Gauss-charge}) using the two-sided Fourier transform, for collision frequencies $\nu = 100$, $200$, and $400$~MeV. We note that our results obtained with the parameter values used by the authors of \cite{Grayson:2022asf} agree with their results. It turns out that the very short time interval during which the magnetic field generated by the point-like charge is significantly stronger than that produced by the Gaussian charge distribution has a substantial and long-lasting impact on the magnitude of the induced currents, as seen in Fig.~\ref{Fig-B-finite-size-medium}. Consequently, calculations based on the point-like-charge approximation overestimate the magnetic field in the plasma medium.

As already mentioned, the calculations using the one-sided Fourier transform are considerably more involved than those using the two-sided transform. Instead of the trivial frequency integration resulting from the delta function $\delta(\omega -{\bf k} \cdot {\bf v})$, one must evaluate the frequency integral using the residue theorem. Even for the Gaussian charge distribution, this requires a numerical approach. We therefore plan to calculate the electromagnetic field using a more realistic Woods-Saxon nuclear charge distribution rather than the Gaussian one, and to present the results in a subsequent publication.

\section{Conclusions and outlook}

Using the one-sided Fourier transformation, we have shown that the currents induced in the quark-gluon plasma, which sustain the magnetic field generated in relativistic heavy-ion collisions, develop with a delay and are consequently less effective than those predicted using the usual two-sided Fourier transformation. Since the plasma initially consists predominantly of electrically neutral gluons, while electrically charged quarks emerge at a later stage, the induced currents are delayed even further. Consequently, for a time of about $0.5$ fm/$c$, the magnetic field remains essentially the same as in vacuum, and the magnitude of the field in the quark-gluon plasma is significantly lower than suggested by earlier estimates. This may explain why magnetic-field effects are difficult to observe in relativistic heavy-ion collisions.

Since the paper is primarily devoted to presenting the one-sided Fourier-transform method and comparing its results with those obtained using the usual two-sided transform, we have adopted several simplifying approximations. More detailed results, not only for the magnetic field but also for the electric field, obtained using a more realistic treatment of the colliding nuclei and the quark-gluon plasma, will be presented in a subsequent publication.

\section*{Acknowledgments}

The collaboration with Patrycja S\l o\'n at an early stage of the project is gratefully acknowledged.

\appendix

\section{One-sided Fourier transform for initial condition at $t_0>0$}
\label{app-initial-t0}

If the initial condition is imposed not at time $t=0$ but at time $t=t_0>0$, the one-sided Fourier transform is modified as
\be
\label{one-sided-Fourier-t0}
f(\omega,{\bf k})=\int_{t_0}^{\infty}{d t}\int{d^3r\, e^{i(\omega t - {\bf k} \cdot {\bf r})}f(t,{\bf r})} .
\ee
Changing the time variable $t \to t - t_0$, one shows that the inverse transformation is not changed and is given by Eq.~(\ref{inv-one-sided-Fourier}).

Applying the transformation (\ref{one-sided-Fourier-t0}) to the Maxwell equations (\ref{Maxwell-eqs-x}), one obtains Eqs.~(\ref{Maxwell-eqs-k}) but instead of $\mathbf{B}_0({\bf k})$ and $\mathbf{D}_0({\bf k})$ there are $e^{i\omega t_0} \mathbf{B}_0(t_0,{\bf k})$ and $e^{i\omega t_0} \mathbf{D}_0(t_0,{\bf k})$, respectively.  The initial fields at $t=0$ given by Eqs.~(\ref{E0-point}) and (\ref{B0-point}), are now replaced by 
\ban
\label{E0-point-t0}
{\bf E}_0 (t_0,{\bf k}) &=& - 4 \pi i \, Ze \, e^{-i {\bf k}\cdot({\bf v} t_0+{\bf r}_0)}
 \frac{({\bf k} \cdot  {\bf v}) {\bf v} - {\bf k}}{({\bf k} \cdot  {\bf v})^2 - {\bf k}^2}  
\, e^{-i {\bf k} \cdot  {\bf r}_0} ,
\\[4mm]
\label{B0-point=t0}
{\bf B}_0(t_0,{\bf k}) &=& - 4 \pi i  \, Ze \, e^{-i {\bf k}\cdot({\bf v} t_0+{\bf r}_0)}
\frac{{\bf k} \times {\bf v}}{({\bf k} \cdot  {\bf v})^2 - {\bf k}^2} 
 \, e^{-i {\bf k} \cdot  {\bf r}_0}.
\ean

\section{Discrete Hankel Transform}
\label{app-Hankel}

The Hankel transform of order $n$ is defined as 
\be
\label{Hankel-def}
\tilde{f}(k) = \int_0^{\infty} dr \, r \, f(r) \, J_n(k \, r) ,
\ee
where $J_n(z)$ is the Bessel function of the first kind of order $n$. The inverse transform is
\be
\label{Hankel-inv}
f(r)=\int_0^{\infty}dk \, k \, \tilde{f}(k) \, J_{n}(k \, r) .
\ee
We use only the transform of order $n=1$ in our magnetic field computations but following \cite{Ronen:2006} we present here the general case. 

Our aim is to numerically compute the integral of the form (\ref{Hankel-inv}) and the key element of the adopted algorithm is the following orthogonality relation of the $n$-th Bessel function
\be
\label{eq:bessel_ortho}
\int_0^1 dx \, x \, J_n(x \, u_\alpha) \, J_n(x \, u_\beta)
=\frac{1}{2} \, \delta^{\alpha\beta} J^2_{n+1}(u_\alpha) ,
\ee
where $u_\alpha$ is the $\alpha$-th zero of the Bessel function of order $n$. To use the formula (\ref{eq:bessel_ortho}) to compute the integral (\ref{Hankel-inv}), it is necessary that the function $f(r)$ vanishes for $r > R_{\rm max}$. Then, the upper limit of the integral (\ref{Hankel-def}) can be set as $R_{\rm max}$. Introducing the variable $r = x R_{\rm max}$, the orthogonality condition is rewritten as
\be
\label{orthogonality}
\int_0^{R_{\rm max}} dr\,r J_n ( r \, k_\alpha) \, J_n (r \, k_\beta)
=\frac{1}{2} \, \delta^{\alpha\beta} R_{\rm max}^2 J^2_{n+1}(u_\alpha) ,
\ee
where $k_\alpha = u_\alpha / R_{\rm max}$ with $\alpha = 1, 2, 3, \dots$ are discrete points in the momentum space. Expressing the function $f(r)$ as a finite series
\be
f(r) = \sum_{\alpha = 1}^{N} c_\alpha J_n (r  \, k_\alpha),
\ee
and using Eqs.~(\ref{Hankel-inv}) and (\ref{orthogonality}), one finds the coefficients $c_\alpha$ as
\be
c_\alpha =\frac{2}{R_{\rm max}^2} \frac{\tilde{f}(k_\alpha)}{J^2_{n+1}(u_\alpha)} .
\ee

Finally, one obtains the integral of interest (\ref{Hankel-inv}) expressed in the desired form
\be
\int_0^{K_{\rm max}} dk \, k \, \tilde{f}(k) \, J_n(r \, k) 
 = \frac{2}{R_{\rm max}^{2}} 
 \sum_{\alpha =1}^{N}{\frac{J_n (r \, k_\alpha)}{J^2_{n+1}(u_\alpha)} \, \tilde{f}(k_\alpha)} ,
\ee
where the upper cut-off $K_{\rm max}=u_{N+1}/R_{\rm max}$ with $u_{N+1}$ being the ($N$+1)-th zero of the Bessel function of order $n$.



\begin{thebibliography}{99}

\bibitem{Tuchin:2013ie}
K.~Tuchin,
Adv. High Energy Phys. \textbf{2013}, 490495 (2013).

\bibitem{Kharzeev:2012ph}
D.~E.~Kharzeev, K.~Landsteiner, A.~Schmitt and H.~U.~Yee,
Lect. Notes Phys. \textbf{871}, 1-11 (2013).

\bibitem{Fukushima:2018grm}
K.~Fukushima,
Prog. Part. Nucl. Phys. \textbf{107}, 167 (2019).

\bibitem{Huang:2015oca}
X.~G.~Huang,
Rept. Prog. Phys. \textbf{79}, 076302 (2016).

\bibitem{STAR:2023bzh}
B.~E.~Aboona \textit{et al.} [STAR],
Phys. Lett. B \textbf{862}, 139245 (2025).

\bibitem{STAR:2023jdd}
M.~I.~Abdulhamid \textit{et al.} [STAR],
Phys. Rev. X \textbf{14}, 011028 (2024).

\bibitem{Wang:2022eoo}
F.~Wang,
Acta Phys. Polon. Supp. \textbf{16}, 1-A15 (2023).

\bibitem{STAR:2025vhs}
B.~E.~Aboona \textit{et al.} [STAR],
Phys. Rev. C \textbf{113}, 014912 (2026).

\bibitem{Shen:2025unr}
D.~Shen, J.~Chen, X.~G.~Huang, Y.~G.~Ma, A.~Tang and G.~Wang,
Research \textbf{8}, 0726 (2025).

\bibitem{Kharzeev:2007jp}
D.~E.~Kharzeev, L.~D.~McLerran and H.~J.~Warringa,
Nucl. Phys. A \textbf{803}, 227 (2008).

\bibitem{Skokov:2009qp}
V.~Skokov, A.~Y.~Illarionov and V.~Toneev,
Int. J. Mod. Phys. A \textbf{24}, 5925 (2009).

\bibitem{Voronyuk:2011jd}
V.~Voronyuk, V.~D.~Toneev, W.~Cassing, E.~L.~Bratkovskaya, V.~P.~Konchakovski and S.~A.~Voloshin,
Phys. Rev. C \textbf{83}, 054911 (2011).

\bibitem{Bzdak:2011yy}
A.~Bzdak and V.~Skokov,
Phys. Lett. B \textbf{710}, 171 (2012).

\bibitem{Tuchin:2010vs}
K.~Tuchin,
Phys. Rev. C \textbf{82}, 034904 (2010)
[erratum: Phys. Rev. C \textbf{83}, 039903 (2011)].

\bibitem{Deng:2012pc}
W.~T.~Deng and X.~G.~Huang,
Phys. Rev. C \textbf{85}, 044907 (2012).

\bibitem{McLerran:2013hla}
L.~McLerran and V.~Skokov,
Nucl. Phys. A \textbf{929}, 184 (2014).

\bibitem{Gursoy:2014aka}
U.~Gursoy, D.~Kharzeev and K.~Rajagopal,
Phys. Rev. C \textbf{89}, 054905 (2014).

\bibitem{Zakharov:2014dia}
B.~G.~Zakharov,
Phys. Lett. B \textbf{737}, 262-266 (2014).

\bibitem{Roy:2015kma}
V.~Roy, S.~Pu, L.~Rezzolla and D.~Rischke,
Phys. Lett. B \textbf{750}, 45 (2015).

\bibitem{Li:2016tel}
H.~Li, X.~l.~Sheng and Q.~Wang,
Phys. Rev. C \textbf{94}, 044903 (2016).

\bibitem{Huang:2022qdn}
A.~Huang, D.~She, S.~Shi, M.~Huang and J.~Liao,
Phys. Rev. C \textbf{107}, 034901 (2023).

\bibitem{Grayson:2022asf}
C.~Grayson, M.~Formanek, J.~Rafelski and B.~Mueller,
Phys. Rev. D \textbf{106}, 014011 (2022).

\bibitem{Yan:2021zjc}
L.~Yan and X.~G.~Huang,
Phys. Rev. D \textbf{107}, 094028 (2023).

\bibitem{Wang:2021oqq}
Z.~Wang, J.~Zhao, C.~Greiner, Z.~Xu and P.~Zhuang,
Phys. Rev. C \textbf{105}, L041901 (2022).

\bibitem{Rukhadze-Silin-1961}
A.~A.~Rukhadze and V.~P.~Silin, 
Usp. Fiz. Nauk {\bf 74}, 223 (1961) (in Russian) [Sov. Phys. Usp. {\bf 4}, 459 (1961)].

\bibitem{LP81} 
E.~M.~Lifshitz and L.~P.~Pitaevskii,
{\it Physical Kinetics} (Pergamon Press, Oxford, 1981).

\bibitem{Le-Bellac-2000}
M.~ Le~Bellac, 
{\it Thermal Field Theory} 
(Cambridge University Press, Cambridge, 2000).

\bibitem{Mrowczynski:1988xu}
St.~Mr\'owczy\'nski,
Acta Phys. Polon. B \textbf{19}, 91 (1988).

\bibitem{Ding:2010ga}
H.~T.~Ding, A.~Francis, O.~Kaczmarek, F.~Karsch, E.~Laermann and W.~Soeldner,
Phys. Rev. D \textbf{83}, 034504 (2011).

\bibitem{Ronen:2006}
S.~Ronen, D.~C.~E.~Bortolotti, J.~L.~Bohn, 
Phys. Rev. A \textbf{74}, 013623 (2006).

\bibitem{DeJager:1974liz}
C.~W.~De Jager, H.~De Vries and C.~De Vries,
Atom. Data Nucl. Data Tabl. \textbf{14}, 479 (1974);
[erratum: Atom. Data Nucl. Data Tabl. \textbf{16}, 580 (1975)].



\end{thebibliography}
\end{document}